\documentclass[aps,prd,preprint,superscriptaddress,showpacs,tightenlines]{revtex4}
\usepackage{graphics}
\usepackage{epsfig}
\usepackage{latexsym}
\usepackage{colordvi}
\usepackage{amsmath}
\usepackage{amssymb}

\newcommand{\nc}{\newcommand}
\nc{\beq}{\begin{equation}}
\nc{\eeq}{\end{equation}}
\nc{\beqa}{\begin{eqnarray}}
\nc{\eeqa}{\end{eqnarray}}

\input epsf
\newwrite\ffile\global\newcount\figno \global\figno=1

\def\writedef#1{}
\def\figin{\epsfcheck\figin}\def\figins{\epsfcheck\figins}
\def\epsfcheck{\ifx\epsfbox\UnDeFiNeD
\message{(NO epsf.tex, FIGURES WILL BE IGNORED)}
\gdef\figin##1{\vskip2in}\gdef\figins##1{\hskip.5in}
\else\message{(FIGURES WILL BE INCLUDED)}%
\gdef\figin##1{##1}\gdef\figins##1{##1}\fi}
\def\figinsert{}
\def\ifig#1#2#3{\xdef#1{fig.~\the\figno}
\writedef{#1\leftbracket fig.\noexpand~\the\figno}%
\figinsert\figin{\centerline{#3}}\medskip\centerline{\vbox{\baselineskip12pt
\advance\hsize by -1truein\center\footnotesize{  Fig.~\the\figno.} #2}}
\bigskip\endinsert\global\advance\figno by1}
\def\endinsert{}

\begin{document}

\preprint{\parbox[b]{1in}{ \hbox{\tt PNUTP-03/A03} \hbox{\tt OITS
730} }}


\title{Positivity and Dense Matter}

\author{Deog Ki Hong}


\affiliation{ Department of Physics, Pusan National University,
Pusan 609-735, Korea
\protect \\{\footnotesize\tt
dkhong@pusan.ac.kr}}

\author{Stephen D.H.~Hsu}
\affiliation{
Department of Physics,
University of Oregon, Eugene OR 97403-5203
\protect\\ {
\footnotesize\tt hsu@duende.uoregon.edu}}

\vspace{0.1in}

\date{\today}

\begin{abstract}
We elaborate on previous results concerning the positivity of the
Euclidean path integral measure for low-energy modes in dense
fermionic matter. We show that the sign problem usually associated
with fermions is absent if one considers only low-energy degrees
of freedom. We describe a method for simulating dense QCD on the
lattice and give a proof using rigorous inequalities that the
color-flavor locked (CFL) phase is the true vacuum of three
flavor, massless QCD. We also discuss applications to electronic
systems in condensed matter, such as generalized Hubbard models.
\end{abstract}

\maketitle


\newpage

\section{Introduction}

Euclidean quantum chromodynamics (QCD) with a non-zero chemical
potential has a complex measure, which has made lattice simulation
particularly difficult~\cite{Hands:2001jn}. (For recent lattice
investigation of the QCD phase boundary at finite density, see
\cite{Fodor:2001au,Allton:2002zi}).
This problem is often referred to as the sign
problem, because by appropriately grouping terms quantities such
as the partition function can be written as a sum over real, but
potentially negative, terms. (That this grouping can be
accomplished is in many systems a consequence of a discrete
symmetry such as parity or time-reversal invariance.)
Indeterminate signs are enough to preclude use of importance
sampling, the main technique for speeding up Monte Carlo
integration. It is important to note that while the sign problem
often arises in systems of fermions, it is neither inevitable nor
inescapable. For example, in QCD at zero chemical potential and in
the Hubbard model at half filling one can organize the sum so that
terms are real and positive. For recent work on the sign problem,
see \cite{sign}.

Analytical work in color superconductivity \cite{csc} has
demonstrated a rich phase structure at high density, and
stimulated interest in QCD at non-zero baryon density. Several
experiments have been proposed to probe matter at density of a few
times nuclear matter density \cite{exper}. Even rudimentary
information about the behavior of dense matter would be useful to
the experimental program, as well as to the study of compact
astrophysical objects such as neutron stars.
In an earlier paper~\cite{Hong:2002nn},
we showed that QCD near a Fermi surface has
positive, semi-definite measure. In the limit of low energies, the
contribution of the remaining modes far from the Fermi surface can
be systematically expanded, using a high density effective theory
previously introduced by one of us~\cite{Hong:2000tn,Hong:2000ru}.
This
effective theory is sufficient to study phenomena like color
superconductivity, although quantities like the equation of state
may be largely determined by dynamics deep in the Fermi sea.

The expansion about the Fermi surface is in powers of $1/ \mu$,
where $\mu$ is the chemical potential. For this expansion to be
controlled, the ultraviolet cutoff of our effective theory must be
less than $\mu$, or equivalently the scale of the physics of
interest must be small relative to the chemical potential. In QCD
at asymptotic density, the superconducting gap is exponentially
small, so this condition is satisfied. However, it is also quite
possible that at intermediate densities (e.g., those inside a
neutron star) the gap is somewhat smaller than $\mu$, providing us
with an additional small dimensionless parameter. Even if this is
not the case, the power expansion of the effective theory is
qualitatively different from the usual perturbation in $\alpha_s$,
and therefore worth exploring. Finally, we note that models (e.g.,
of electronic systems) which are not asymptotically free may
exhibit strongly coupled quasiparticle excitations even at high
density. The results described here still apply to such systems
and may be of use in their simulation.

This paper is organized as follows. We begin in section 2 with
simple examples in (1+1) and higher dimensions which illustrate
how the effective Fermi surface description can have positive
measure even if the original model has a sign problem. In section
3 we review the results from our previous paper, and include some
discussion of how they apply to electronic systems that may arise
in condensed matter. In section 4 we describe how they can be
applied to lattice simulations of dense matter. In section 5 we
discuss correlator inequalities (also known as QCD inequalities)
that result from positivity, and how they restrict the possible
ground states of QCD. In section 6 we conclude with a summary and
future prospects.

\section{Example: (1+1) Dimensions and Beyond}

We begin with an example that illustrates the basic ideas in a simple
setting. Consider the Euclidean (1+1) action of non-relativistic
fermions interacting with a gauge field A
\begin{equation}
\label{NRA} S = \int d\tau dx ~\psi^*_\sigma \left[ (
- \partial_\tau + i\phi + \epsilon_F )  ~-~ \epsilon( -i\partial_x +
A ) \right] \psi_\sigma
\end{equation}
where $\epsilon (p)$ is the energy as a function of momentum (e.g.
$\epsilon (p) \approx {\frac{p^2}{2m}} + \cdots$). In (\ref{NRA})
and below we may consider it as a function of the operator
($-i\partial_x + A$). The dispersion relation in the presence of
the chemical potential $\epsilon_F$ is: $E(p) = \epsilon(p) -
\epsilon_F$, and a low energy mode must have momentum close to
$\pm p_F$, where $\epsilon ( \pm p_F ) = \epsilon_F$. The Fermi
surface in (1+1) dimensions is reduced to the two points $p = \pm
p_F$. Near these points we have
\begin{equation}
E( p \pm p_F ) \approx \pm~ v_F p~~~,
\end{equation}
where $v_F = \partial E / \partial p \vert_{p_F}$ is the Fermi
velocity.

The action (\ref{NRA}) is not obviously positive. In fact, the
operator in brackets $\left[ ~\cdots ~\right]$ clearly has
Hermitian as well as anti-Hermitian components, and hence complex
eigenvalues.

Let us assume that the gauge field has small amplitude and is
slowly varying relative to the scale $p_F$. We will extract the
slowly varying component of the fermion field to construct a low
energy effective theory involving quasiparticles and gauge fields.
This effective theory will have positive, semi-definite
determinant.

First, we extract the quasiparticle modes (we suppress the spin
index in what follows)
\begin{equation}
\psi (x, \tau) = \psi_L e^{+i p_F x} ~+~\psi_R e^{-i p_F x}~~~,
\end{equation}
where the functions $\psi_{L,R}$ are slowly varying. To simplify
the action, we use the identity
\begin{equation}
e^{\pm ip_F x} ~E( -i\partial_x + A )~ e^{\mp ip_F x} ~\psi (x)
\approx \pm~ v_F (-i\partial_x + A) \psi(x)~~~,
\end{equation}
to obtain~\cite{vf}
\begin{equation}
\label{EFA} S_{\rm eff} = \int d\tau dx\left[\psi^{\dagger}_L ( - \partial_\tau
+ i\phi + i\partial_x - A ) \psi_L  ~+~ \psi^*_R ( - \partial_\tau +
i\phi - i\partial_x + A ) \psi_R\right].
\end{equation}
We can write this in a more familiar form by introducing the
Euclidean (1+1) gamma matrices $\gamma_{0,1,2}$~, which are
Hermitian and can be taken as $\gamma_i = \sigma_i$ where
$\vec{\sigma}$ are the Pauli matrices.
Using $\psi_{L,R} = \frac{1}{2} ( 1 \pm \gamma_2) \psi$ we obtain
\begin{equation}
\label{SReff} S_{\rm eff} = \int d\tau dx ~  \bar{\psi}
\gamma^{\mu} (\partial_{\mu} + iA_{\mu} ) \psi ~\equiv~ \int d\tau
dx ~ \bar{\psi} D\!\!\!\!/ \psi ~~~.
\end{equation}
Since the gamma matrices are Hermitian, and the operator
$(\partial_{\mu} + iA_{\mu} )$ is anti-Hermitian, the operator
$D\!\!\!\!/$ in (\ref{SReff}) has purely imaginary eigenvalues.
However, because $\gamma_2$ anticommutes with $D\!\!\!\!/$, the
eigenvalues come in conjugate pairs: given $D\!\!\!\!/~ \phi =
\lambda \phi$, we have
$$D\!\!\!\!/ ~(\gamma_2 \phi) = - \gamma_2 D\!\!\!\!/ ~\phi = -
\gamma_2 \lambda \phi = - \lambda (\gamma_2 \phi_n)~~~.$$ Hence
the determinant $\det D\!\!\!\!/ ~= \prod \lambda^* \lambda$ is
real and positive semi-definite.

Thus, by considering only the low-energy modes near the Fermi
points of the original model (\ref{NRA}), we obtain an effective
theory with desirable positivity properties. Note that it is
necessary that the interactions (in this case, the background
gauge field A) not couple strongly the low-energy modes to fast modes which
are far from the Fermi points. This is a reasonable approximation
in many physical situations, where it is the interactions among
quasiparticles that are of primary interest. In what follows, we
will apply this basic idea to more complex models such as QCD.

It is straightforward to go beyond (1+1) dimensions. Consider an
electron system, described by
\begin{equation}
{\cal L}=\psi^{\dagger}\left[i\partial_t-{\epsilon(\vec p)}\right]\psi
+\mu\psi^{\dagger}\psi,
\end{equation}
where  $\epsilon(\vec p)$ is the electron energy, a function of
momentum $\vec p$.
It is interesting to note that the non-relativistic system already
has a sign problem even at the zero density, $\mu=0$, though the
free case does not suffer this, thanks to  the
separation of variables. In fact, it is quite unusual
to have a system like vacuum QCD which has no sign problem.
In Euclidean space the electron determinant is
\begin{equation}
\label{edet}
M=-\partial_{\tau}-\epsilon (\vec p)+\mu.
\end{equation}
The first term in operator (\ref{edet}) is anti-Hermitian, while the rest are
Hermitian. Since there is no constant matrix $P$ in the spin space
that satisfies $M^{\dagger}=P \, M\,P^{-1}$, it has a sign problem
in general.

Let us decompose the fermion momentum as
\begin{equation}
\vec p=\vec p_F+\vec l.
\end{equation}
Again, the Fermi momentum is defined to be a momentum at which the energy
equals to the chemical potential at zero temperature:
$\mu=\epsilon(p_F)$, and the Fermi velocity is defined as
\begin{eqnarray}
\vec v_F=\left.\frac{\partial\epsilon(p)}{\partial{\vec p}}\right|_{p=p_F}.
\end{eqnarray}
If we are interested in low energies, $\left|\vec l\right|\ll p_F$,
we may integrate out the fast modes to get an effective operator,
\begin{equation}
M_{\rm EFT}=-\partial_{\tau}-\vec v_F\cdot \vec l,
\end{equation}
which has  complex eigenvalues. However, when we include the
$-\vec v_F$ sector, we have $M_{\rm EFT}(\vec v_F)M_{\rm EFT}(-\vec v_F)\le
0$ (i.e., has real negative eigenvalues),
assuming $\epsilon(\vec p)=\epsilon(-\vec p)$.
We again see that the sign problem in the electron system is alleviated in the
low-energy effective theory.

\section{QCD}

Let us recall why the measure of dense QCD is complex in Euclidean
space. We use the following analytic continuation of the Dirac
Lagrangian to Euclidean space:
\begin{equation}
 x_0 \rightarrow -i x_E^4,\quad x_i \rightarrow x_E^i
~;~ \gamma_0 \rightarrow \gamma_E^4,\quad \gamma_i \rightarrow
i\gamma_E^i ~~~.
\end{equation}
The Euclidean gamma matrices satisfy
\begin{equation}
{\gamma_E^{\mu}}^{\dagger}=\gamma_E^{\mu} ~~,~~
\left\{\gamma_E^{\mu},\gamma_E^{\nu}\right\}=2\delta^{\mu\nu}.
\end{equation}
The Dirac-conjugated field, $\bar\psi=\psi^{\dagger}\gamma^0$, is
mapped into a field, still denoted as $\bar\psi$, which is
independent of $\psi$ and transforms as $\psi^{\dagger}$ under
$SO(4)$. Then, the grand canonical partition function for QCD is
\begin{eqnarray}
Z(\mu)=\int {\rm d}A_{\mu}\det \left(M\right)e^{-S(A_{\mu})},
\end{eqnarray}
where $S(A_{\mu})$ is the positive semi-definite gauge action, and
the Dirac operator
\begin{equation}
\label{M} M=\gamma_E^{\mu}D_E^{\mu}+\mu\gamma_E^4,
\end{equation}
where $D_E = \partial_E + iA_E$ is the analytic continuation of
the covariant derivative. The Hermitian conjugate of the Dirac
operator is
\begin{equation}
M^{\dagger}=-\gamma_E^{\mu}D_E^{\mu}+\mu\gamma_E^4~~~.
\end{equation}
The first term in (\ref{M}) is anti-Hermitian, while the second is
Hermitian, hence the generally complex eigenvalues. When $\mu =
0$, the eigenvalues are purely imaginary, but come in conjugate
pairs $(\lambda, \lambda^*)$~\cite{ev}, so the resulting
determinant is real and positive semi-definite:
\begin{equation}
\rm det~ M =\prod \lambda^*\lambda \ge0~~~.
\end{equation}

In what follows we investigate the positivity properties of an
effective theory describing only modes near the Fermi surface. A
system of degenerate quarks with a net baryon number asymmetry is
described by the QCD Lagrangian density with a chemical potential
$\mu$,
\begin{equation}
{\cal L}_{\rm QCD}=\bar\psi i D\!\!\!\!/ ~\psi
-\frac{1}{4}F_{\mu\nu}^aF^{a\mu\nu}+\mu \bar\psi\gamma_0\psi,
\label{lag}
\end{equation}
where the covariant derivative $D_{\mu}=\partial_{\mu}+i A_{\mu}$
and we neglect the quark mass for simplicity.

At high density ($\mu\gg\Lambda_{\rm QCD}$),
due to asymptotic freedom the energy spectrum of quarks
near the Fermi surface is  approximately
given  by a free Dirac eigenvalue equation,
\begin{equation}
\left(\vec\alpha\cdot \vec
p-\mu\right)\psi_{\pm}=E_{\pm}\psi_{\pm},
\end{equation}
where $\vec\alpha=\gamma_0\vec\gamma$ and $\psi_{\pm}$ denote the
energy eigenfunctions with eigenvalues $E_{\pm}=-\mu\pm \left|\vec
p\right|$, respectively. At low energy $E<\mu$, the states
$\psi_+$ near the Fermi surface,  $|\vec p|\sim\mu$, are easily
excited but $\psi_-$, which correspond to the states in the Dirac
sea, are completely decoupled due to the presence of the energy
gap $\mu$ provided by the Fermi sea. Therefore the appropriate
degrees of freedom at low energy consist of gluons and $\psi_+$
only.

Now, we wish to construct an effective theory describing the
dynamics of $\psi_+$ by integrating out modes whose energy is
greater than $\mu$. Consider a quark near the Fermi surface, whose
momentum is close to $\mu\vec v_F$. Without loss of generality, we
may decompose the momentum of a quark into a Fermi momentum and a
residual momentum as
\begin{equation}
\label{decomp} p_{\mu}=\mu v_{\mu}+l_{\mu},
\end{equation}
where $v^{\mu}=(0,\vec v_F)$. Since the quark energy is given as
\begin{equation}
E=-\mu+\sqrt{(l_{\parallel}+\mu)^2+l_{\perp}^2},
\end{equation}
the residual momentum should satisfy
$(l_{\parallel}+\mu)^2+l_{\perp}^2\le4\mu^2$ with $\vec
l_{\parallel}=\vec v_F\vec l\cdot \vec v_F$ and $\vec
l_{\perp}=\vec l-\vec l_{\parallel}$.

To describe the small excitations of the quark with Fermi
momentum, $\mu\vec v_F$, we decompose the quark fields as
\begin{equation}
\label{psidecomp} \psi(x)=e^{i\mu\vec v_F\cdot \vec x}\left[
\psi_+(\vec v_F,x)+ \psi_-(\vec v_F,x)\right],
\end{equation}
where
\begin{equation}
\psi_{\pm}(\vec v_F,x)=P_{\pm}(\vec v_F) e^{-i\mu\vec v_F\cdot\vec
x}\psi(x) \quad{\rm with}\quad P_{\pm}(\vec v_F)\equiv\frac{1\pm \vec
\alpha\cdot\vec v_F}{2}.
\end{equation}
The quark Lagrangian in Eq.~(\ref{lag}) then becomes
\begin{eqnarray}
\label{expand} \bar\psi \left(i D\!\!\!\!/
+\mu\gamma^0\right)\psi&=&\left[\bar\psi_+(\vec
v_F,x)i\gamma^{\mu}_{\parallel}D_{\mu}\psi_+(\vec v_F,x)
+\bar\psi_-(\vec v_F,x)\gamma^{0}\left(2\mu+i\bar
D_{\parallel}\right)\psi_-(\vec
v_F,x)\right]\nonumber\\
& &+\left[\bar\psi_-(\vec v_F,x)i {D\!\!\!\!/}_{\perp}\psi_+(\vec
v_F,x)+{\rm h.c.}\right]
\end{eqnarray}
where $\gamma^{\mu}_{\parallel}\equiv(\gamma^0,\vec v_F\vec
v_F\cdot\vec \gamma)$, $\gamma^{\mu}_{\perp}=\gamma^{\mu}-
\gamma^{\mu}_{\parallel}$, $\bar D_{\parallel}=\bar
V^{\mu}D_{\mu}$ with $V^{\mu}=(1,\vec v_F)$, $\bar
V^{\mu}=(1,-\vec v_F)$, and ${D\!\!\!\!/}
_{\perp}=\gamma^{\mu}_{\perp}D_{\mu}$.

At low energy, we integrate out all the ``fast'' modes $\psi_-$
and derive the low energy effective Lagrangian by matching all the
one-light-particle irreducible amplitudes containing gluons and
$\psi_+$ in loop expansion. The effects of fast modes will appear
in the quantum corrections to the couplings of low energy
interactions. At tree-level, the matching is equivalent to
eliminating $\psi_-$ in terms of equations of motion:
\begin{equation}
\label{eliminate} \psi_-(\vec v_F,x)=-\frac{i\gamma^0}{2\mu
+iD_{\parallel}}{D\!\!\!\!/}_{\perp}\psi_+(\vec v_F,x)=-
\frac{i\gamma^0}{2\mu}
\sum_{n=0}^{\infty}\left(-\frac{iD_{\parallel}}{2\mu}\right)^n
{D\!\!\!\!/}_{\perp}\psi_+(\vec v_F,x). 
\end{equation}
Therefore, the tree-level Lagrangian for $\psi_+$ becomes
\begin{equation}
\label{treeL} {\cal L}_{\rm eff}^0=
\bar\psi_+i\gamma_{\parallel}^{\mu}D_{\mu}\psi_+-\frac{1}{2\mu}\bar\psi_+
\gamma^0({D\!\!\!\!/}_{\perp})^2\psi_+ ~+~ \cdots,
\end{equation}
where the ellipsis denotes terms with higher derivatives.

Consider the first term in our effective Lagrangian, which when
continued to Euclidean space yields the operator
\begin{equation}
M_{\rm eft}=\gamma^{E}_{\parallel}\cdot D(A).
\end{equation}
$M_{\rm eft}$ is anti-Hermitian and it anti-commutes with
$\gamma_5$, so it leads to a positive semi-definite determinant.
However, note that the Dirac operator is not well defined in the
space of $\psi_+(\vec v_F,x)$ (for fixed $v_F$), since it maps
$\psi_+(\vec v_F,x)$ into $\psi_+(-\vec v_F,x)$:
\begin{equation}
i
D_{\parallel}\!\!\!\!\!\!/~~P_+\psi=P_-iD_{\parallel}\!\!\!\!\!\!/~~\psi.
\end{equation}
Since $P_{-}(\vec v_F)=P_{+}(-\vec v_F)$, $iD\!\!\!\!/~\psi_+(\vec
v_F,x)$ are $\psi_+(-\vec v_F,x)$ modes, or fluctuations of a
quark with momentum $-\mu \vec v_F$.

We can demonstrate the necessity of including both $\psi_+(\vec
v_F,x)$ and $\psi_+(-\vec v_F,x)$ modes in our effective theory by
considering charge conservation in a world with only $+ \vec{v}_F$
quarks. The divergence of the quark current at one loop is
at the leading order in the $1/\mu$ expansion
\begin{equation}
\left<\partial_{\mu}J^{a\mu}(\vec v_F,x)\right> =g_s\int
\frac{{\rm d}^4p}{(2\pi)^4}e^{-ip\cdot x}p^{\mu}\Pi^{ab}_{\mu\nu}(p)
A_{\parallel}^{b\nu}(-p)\,,
\end{equation}
where $A_{\parallel}=(A_0,\vec v_F\vec v_F\cdot\vec A)$ and
$\Pi^{ab}_{\mu\nu}$ is the vacuum polarization tensor in the
effective theory given as~\cite{Hong:2000tn}
\begin{equation}
\Pi^{\mu\nu}_{ab}(p) =-i\frac{\mu^2}{2}\delta_{ab}\,
\left( \frac{-2\vec p\cdot\vec
v_FV^{\mu}V^{\nu}}{p\cdot V +i\epsilon \vec p\cdot\vec v_F}\,
+g^{\mu\nu}_{\perp}\right),
\end{equation}
where $g^{\mu\nu}_{\perp}=g^{\mu\nu}-(V^{\mu}{\bar V}^{\nu}+
{\bar V}^{\mu}{ V}^{\nu})/2$ is the metric tensor perpendicular to
$V^{\mu}$ and ${\bar V}^{\mu}$.
The polarization tensor has to be transverse to maintain gauge
invariance. We find that if we have both fields $\psi_+(\vec
v_F,x)$ and $\psi_+(-\vec v_F,x)$ the current is conserved and the
gauge symmetry is not anomalous:
\begin{equation}
\left<\partial_{\mu}J_a^{\mu}(\vec v_F,x)+
\partial_{\mu}J_a^{\mu}(-\vec v_F,x)\right>=0.
\end{equation}
Therefore, we need to introduce quark fields with opposite
momenta. The Dirac operator is well defined on this larger space.

This anomaly can be understood in terms of spectral flow, since
the Fermi surface is (in a certain sense) not gauge-invariant.
Under a gauge transformation, $U(x) = e^{i\vec q\cdot\vec x}$, the
Hamiltonian changes and the energy spectrum of free modes of
residual momentum $\vec{l}$ shifts to $E=\vec l\cdot \vec v_F+\vec
q\cdot\vec v_F$. Quarks near the Fermi surface with $\vec{v}_F
\cdot \vec{q} > 0$ flow out of the Fermi sea, creating charge.
This charge creation is compensated by quarks with opposite
$\vec{v}_F$; their energy decreases and they flow into the Fermi
sea. However, unless modes with opposite velocities (i.e. both
sides of the Fermi sphere) are included, charge is not conserved.

Thus far we have considered the quark velocity as a parameter
labelling different sectors of the quark field. This is similar to
the approach of heavy quark effective theory
(HQET)~\cite{Isgur:vq}, in which the velocity of the heavy charm
or bottom quark is almost conserved due to the hierarchy of scales
between the heavy quark mass and the QCD scale. However, this
approach contains an ambiguity often referred to as
``reparameterization invariance'', related to the non-uniqueness
of the decomposition (\ref{decomp}) of quark momenta into a large
and residual component. In the dense QCD case, two $\psi (v_F, x)$
modes whose values of $v_F$ are not very different may actually
represent the same degrees of freedom of the original quark field.
In what follows we give a different formulation which describes
{\it all} velocity modes of the quark field, and is suitable for
defining the quasiparticle determinant.

First, a more precise definition of the breakup of the quark field
into Fermi surface modes. Using the momentum operator in a
position eigenstate basis: $\vec{p} = -i \vec{\partial}$, we
construct the Fermi velocity operator:
\begin{equation}
\label{velo} \vec{v} =   \frac{-i }{\sqrt{- \nabla^2}}
~\frac{\partial}{\partial \vec{x}}~~,
\end{equation}
which is Hermitian, and a unit vector.

Using the velocity operator, we define the projection operators
$P_\pm$ as before and break up the quark field as, $\psi(x) =
\psi_+ (x) + \psi_- (x)$, with $\psi_\pm = P_\pm \psi$. By leaving
$\vec{v}$ as an operator we can work in coordinate space without
introducing the HQET-inspired velocity Fourier transform which
introduces $v_F$ as a parameter. If we expand the quark field in
the eigenstates of the velocity operators, we recover the previous
formalism with all Fermi velocities summed up.

The leading low-energy part of the quark action is given by
\begin{equation}
\label{leading} {\cal L}_+ =  \bar{\psi} P_- (v) \left( i
\partial\!\!\!/ - A\!\!\!/ + \mu \gamma_0
\right) P_+ (v) \psi~~.
\end{equation}
As before, we define the fields $\psi_+$ to absorb the large Fermi
momentum:
\begin{equation}
\label{absorb}
 \psi_+ (x) = e^{- i \mu \vec{x} \cdot \vec{v} } P_+
(v) \psi(x).
\end{equation}
Let us denote the eigenvalue $v$ obtained by acting on the field
$\psi$ (which has momentum of order $\mu$) as $v_l$ (or $v$
``large''), whereas eigenvalues obtained by acting on the
effective field theory modes $\psi_+$ are denoted $v_r$ (or $v$
``residual''). If the original quark mode had momentum p with $|p|
> \mu$ (i.e. was a particle), then $v_l$ and $v_r$ are parallel,
whereas if $|p| < \mu$ (as for a hole) then $v_r$ and $v_l$ are
anti-parallel. In the first case, we have $P_+ ( v_l ) = P_+
(v_r)$ whereas in the second case $P_+ (v_l) = P_- (v_r)$. Thus,
the residual modes $\psi_+$ can satisfy either of $P_\pm (v_r)
\psi_+ = \psi_+$, depending on whether the original $\psi$ mode
from which it was derived was a particle or a hole. In fact,
$\psi_+$ modes can also satisfy either of $P_\pm (v_l) \psi_+ =
\psi_+$ since they can originate from $\psi$ modes with momentum
$\sim + \mu v$ as well as $- \mu v$ (both are present in the
original measure: $D \bar{\psi} \, D \psi$). So, the functional
measure for $\psi_+$ modes contains all possible spinor functions
-- the only restriction is on the momenta: $|l_0|, |\vec{l}| <
\Lambda$, where $\Lambda$ is the cutoff.

In light of the ambiguity between $v_l$ and $v_r$, the equation
$\psi = e^{+ i \mu x \cdot v } \psi_+$ must be modified to
\begin{equation}
\psi = \exp \left( + i \mu x \cdot v ~ \alpha \cdot v \right)
\psi_+ = \exp \left( + i \mu x \cdot v_r ~\alpha \cdot v_r \right)
\psi_+ ~~~,
\end{equation}
where the factor of $\alpha \cdot v_r$ corrects the sign in the
momentum shift if $v_r$ and $v_l$ are anti-parallel. In general,
any expression with two powers of $v$ is unaffected by this
ambiguity. For notational simplicity we define a local operator
\begin{equation}
X ~\equiv~ \mu ~x\cdot v ~ \alpha \cdot v ~=~ \mu\frac{\alpha^i
x^j}{ \nabla^2}\frac{\partial^2}{\partial x^i
\partial x^j}.
\end{equation}

Taking this into account, we obtain the following action:
\begin{equation}
\label{leading1} {\cal L}_+ = \bar{\psi}_+ e^{- i X} \left( i
\partial\!\!\!/ - A\!\!\!/ + \mu \gamma_0
\right) e^{+ i X} \psi_+ ~~.
\end{equation}
We treat the $A\!\!\!/$ term separately from $i
\partial\!\!\!/ + \mu \gamma_0$ since the former does not commute
with X, while the latter does. Continuing to Euclidean space, and
using the identity $P_- \gamma_\mu P_+ = \gamma_\mu^\parallel
P_+$, we obtain
\begin{equation}
\label{leading2} {\cal L}_+ =  \bar{\psi}_+  \gamma^\mu_\parallel
\left(
\partial^\mu + i A^\mu_+ \right) \psi_+ ~~,
\end{equation}
where
\begin{equation}
A^\mu_+ = e^{-iX} ~ A^\mu ~e^{+iX}~~~,
\end{equation}
and all $\gamma$ matrices are Euclidean. The term containing $A$
cannot be fully simplified because $[v,A] \neq 0$. Physically,
this is because the gauge field carries momentum and can deflect
the quark velocity. The redefined $\psi_+$ modes are functions
only of the residual momenta l, and the exponential factors in the
A term reflect the fact that the gluon originally couples to the
quark field $\psi$, not the residual mode $\psi_+$.

The kinetic term in (\ref{leading2}) can be simplified to
\begin{equation}
\gamma^\mu_{\parallel} \partial^\mu = \gamma^\mu \partial^\mu
\end{equation}
since $v \cdot \partial \, v \cdot \gamma = \partial \cdot
\gamma~.$ The action (\ref{leading2}) is the most general
dimension 4 term with the rotational, gauge
invariance~\cite{gauge} and projection properties
appropriate to quark quasiparticles. Therefore, it is a general
consequence of any Fermi liquid description of quark-like
excitations.

The operator in (\ref{leading2}) is anti-Hermitian and leads to a
positive, semi-definite determinant since it anti-commutes with
$\gamma_5$. The corrections given in (\ref{treeL}) are all
Hermitian, so higher orders in the $1/ \mu$ expansion may
re-introduce complexity. The structure of the leading term plus
corrections is anti-Hermitian plus Hermitian, just as in the
original QCD Dirac Lagrangian with chemical potential.

By integrating out the fast modes, the Euclidean QCD partition
function can be rewritten as
\begin{eqnarray}
Z(\mu)=\int {\rm d}A_+~\det M_{\rm eff}(A_+)e^{-S_{\rm eff}(A_+)}.
\end{eqnarray}
The leading
terms in the effective action for gluons (these terms are
generated when we match our effective theory, with energy cutoff
$\Lambda$, to QCD) also contribute only real, positive terms to
the partition function:
\begin{equation}
S_{\rm eff}(A)=\int{\rm
d}^4x_E\left(\frac{1}{4}F_{\mu\nu}^aF_{\mu\nu}^a +\frac{M^2}{
16\pi}\sum_{\,\vec v_F}A_{\perp\mu}^{a}A_{\perp\mu}^{a}\right)
\ge0,
\end{equation}
where $A_{\perp}=A-A_{\parallel}$ and the Debye screening mass is
$M=\sqrt{N_f/(2\pi^2)}g_s\mu$\,. Note that Landau damping is due
to softer quark modes which have not been integrated out, and
therefore do not contribute to matching.

Although the HDET only describes low-energy modes, it still
contains Cooper pairing interactions. This is because Cooper
pairing, in which the quasiparticles have nearly
equal and opposite momenta, is induced by gluonic interactions
with small energy and momentum transfer. That is, although a gluon
exchange (or other interaction) which causes a large angular
deflection of a quasiparticle
$$ \vert \vec{p} \rangle \rightarrow \vert \vec{p}\,'\rangle $$
must involve a large momentum transfer, and hence is not part of
the effective theory, a Cooper pairing interaction
$$
 \vert \vec{p}, -\vec{p} \rangle \rightarrow \vert \vec{p}\,',- \vec{p}\,' \rangle
$$
only involves a small energy and momentum transfer, even if the angle
between $\vec{p}$ and $\vec{p}\,'$ is large. Hence, it is
described by the leading order interaction between soft gluons and
quarks in the effective theory (\ref{leading2}).

Matching of hard gluon effects also leads to four-quark operators
in the effective theory. The addition of these four-quark
operators still leads to a positive action for attractive
channels, since they arise from quasiparticle-gluon interactions
which are originally positive. A simple way to study the
positivity of four-quark operators is to replace them by a vector
field with trivial quadratic term $V_\mu^2$ which couples to
quarks like the original gluon: $V_\mu \bar{\psi} \gamma^\mu
\psi$. Completing the square, we see that the resulting path
integral is positive. Note that this argument does not apply to
interactions involving six or more quarks, or interactions
involving virtual anti-quarks. However, these are always
suppressed by additional powers of $\mu$.

Finally, some comments on more general models than QCD -- in
particular electronic systems in condensed matter -- which are
also affected by a sign problem. We note that as long as the
important dynamics of the system involve energies which are small
compared to the Fermi energy (or are exhibited at a temperature
small compared to the Fermi temperature), the Euclidean
description of the system is likely to have a positive effective
action after modes far from the Fermi surface are integrated out.
Models involving, e.g., attractive four-fermion interactions or
long range gauge fields fall into this category. As long as the
important interactions are soft interactions between quasiparticle
degrees of freedom not too far from the Fermi surface, the model
can be simulated without a sign problem, even if it is strongly
coupled.

As a specific example, consider the Hamiltonian
\begin{equation}
\label{HH}
H = \sum_{ab} t_{ab} c^{\dagger}_a c_b  ~+~ \sum_{abcd} v_{abcd}~
c^{\dagger}_a c_b c^{\dagger}_c c_d ~,
\end{equation} where $\rm abcd$ denote lattice site and spin indices.
This Hamiltonian includes the Hubbard model as a special case. A
constant diagonal part of the kinetic term acts as a chemical
potential $\mu = t_{aa}$, but $t$ may also contain hopping terms
such as a nearest-neighbor off-diagonal term in the tight binding
approximation. The usual kinetic term, taken by itself, produces a
band structure which in momentum space has the form
\begin{equation}
\sum_k E(k)~ c^\dagger_k c_k~~~,
\end{equation}
for some function $E(k)$. A non-zero chemical potential leads to
filling of levels up to some momentum $k_F$. There is an effective
theory description of modes near this surface if the Fourier
transform of the interaction term $v_{abcd}$ only contains soft
interactions (i.e. support for typical momentum transfers less
than $k_F$.)

To obtain the partition function Z we must extend the fields
from the d-dimensional lattice to (d+1)-dimensional Euclidean space, with
finite extent $\beta$ in the time direction.
The action density is given by
\begin{equation}
- \frac{\partial}{\partial \tau} - {\cal H}
\end{equation}
where $\cal H$ is the Hamiltonian density.
Using the Hubbard-Stratonovich transformation, we can rewrite the
interaction term in terms of a functional integral over a
background field U.
\begin{equation}
Z = \frac{1}{N} \int \prod_{ab} dU_{ab} ~e^{ \sum_{cdef} U_{cd}
v^{-1}_{cdef} U_{ef}} ~\det \left[ ~T - U~ \right]~~~,
\end{equation}
where $T = \frac{\partial}{\partial \tau} - t$ contains the time
derivative as
well as the kinetic term t from the Hamiltonian (\ref{HH}).
We assume that the interaction $v$ is real and negative
definite so that the exponential is real and the integral
converges.

The determinant can be expanded about our effective theory. The
leading term is positive and has no sign problem. The correction
terms are suppressed as long as the background field U has support
on momentum scales small compared to $k_F$. (See discussion in
next section, especially equation (\ref{lndet}) with U playing the
role of the gauge field). Typical fields U in the integral are
determined by the exponential term -- if the Fourier transform of
$v$ has little support at large momenta, then the exponent
$\sum_{k}~U(k)~ v^{-1} (k)~ U(k)$ will be large and negative, and
the corresponding modes of $U$ suppressed.

Unfortunately, the Hubbard model itself contains the interaction
term
\begin{equation}
V \sum_i ~n_{i \uparrow} n_{i \downarrow}
\end{equation}
where $n_i = c^\dagger_i c_i$ is the number operator. This
interaction allows all momentum transfers up to $\pi / a$, where
$a$ is the lattice spacing. Because it couples low-energy modes to
fast modes far from the Fermi surface, the resulting partition
function isn't well-approximated by the positive EFT part.
However, a model with modified (softer) interactions would be
approximately positive and might in fact be more physically
realistic.

\section{Lattice Simulation}

The goal of this section is to give a method for simulating QCD at
finite density. We will consider a chemical potential $\mu$ much
larger than $\Lambda_{\rm QCD}$ throughout, and divide the
functional integral over quark excitations into two parts: (I)
modes within a shell of width $\Lambda$ of the Fermi surface, and
(II) modes which are further than $\Lambda$ from the Fermi
surface. We will assume the hierarchy
\begin{equation}
\label{hierarchy} \mu >> \Lambda >> \Lambda_{\rm QCD}~~~.
\end{equation}

The quark determinant in region (I) is well approximated by the
determinant of the leading operator in high density effective
theory (HDET) as long as the first inequality in (\ref{hierarchy})
is satisfied. As discussed in the previous section, it is positive
and real.

Here we will show that the contributions to the effective action
for the gauge field from quark modes in region (II) are small and
vanish as the $\Lambda$ grows large compared to $\Lambda_{\rm
QCD}$.

First consider the theory in Minkowski space. The Dirac operator
is
\begin{equation}
M = i D\!\!\!\!/ + \mu \gamma_0~~~
\end{equation}
and the Dirac equation can be written as
\begin{equation}
i \partial_0 \psi = H \psi~~~
\end{equation}
with
\begin{equation}
H = i \alpha \cdot \partial - \mu~~~
\end{equation}
a Hermitian operator. The break up into regions (I) and (II)
proceeds naturally in terms of energy eigenvalues of H (or $l_0$
in the HDET notation). The low-lying modes in region (I) are
particle states with spatial momenta satisfying $| \vec{p} |
\approx \mu$.

The analytic continuation of region (I) to Euclidean space leads
to the HDET determinant considered previously.

Modes in region (II) all have large energy eigenvalues, at least
as large as $\Lambda$. In considering their effect on physics at
the scale $\Lambda_{\rm QCD}$, we can integrate them out in favor
of local operators suppressed by powers of $\Lambda_{\rm QCD} /
\Lambda$.

To make this concrete, consider the effective action for gauge
fields with field strengths $F_{\mu \nu}$ of order $\Lambda_{\rm
QCD}$. The quark contribution to this effective action is simply
the logarithm of the determinant we wish to compute. It can be
expanded diagrammatically in graphs with external gauge field
lines connected to a single quark loop. Restricting to region
(II), we require that the quark modes in the loop have large H
eigenvalues. Evaluating such graphs leads only to operators which
are local in the external fields $A_\mu (x)$.

The resulting renormalizable (dimension 4) operator is the finite
density equivalent of $F_{\mu \nu}^2$, except that due to the
breaking of Lorentz invariance it contains separate time- and
space-like components which represent the contribution of
high-energy modes to the renormalization of the coupling constant,
and Debye screening. These effects do not introduce a complex
component when continued to Euclidean space.

Higher dimension operators, which involve additional powers of
$F_{\mu \nu}$ or covariant derivatives $D_\mu$ are suppressed by
the scale $\Lambda$. These are presumably the source of complex
terms introduced to the effective action. However, due to the
$1/\Lambda$ suppression they are dominated by the contribution
from the low-lying modes in region (I), which is necessarily
non-local, but real.

The logarithm of the Euclidean quark determinant will have the
form:
\begin{equation}
\label{lndet} {\rm ln \, det \,} M ~\sim~~ {\cal O}(\mu^4) ~+~ (
{\rm non-local, real}) ~+~ {\cal O}( \frac{1}{\Lambda})({\rm
local, complex})~~~,
\end{equation}
where the first term is the (real, constant) free energy of
non-interacting quarks, the second term is from the positive
determinant in region (I) and the last term is the suppressed,
complex contribution from region (II). Only the last two terms
depend on the gauge field $A_\mu (x)$

On the lattice, one can use the dominant dependence of ${\rm det}
M$ on the first and second terms to do importance sampling. In
order to keep the complex higher dimension operators (last term in
(\ref{lndet})) small, it is important that the gauge field
strengths are kept smaller than $\Lambda^2$. One can impose this
condition by using two different lattice spacings, $a_{g}$ for the
gluons and  $a_{\det}$ for the quarks, with  $a_{g} > a_{\det}$.
The determinant is calculated on the finer $a_{\det}$ lattice, and
is a function of plaquettes which are obtained by interpolation
from the plaquettes on the coarser $a_{g}$ lattice. Interpolation
can be defined in a natural way, since each lattice link variable
$U_{x \mu}$ is an element of the gauge group, and one can connect
any two points $g_1, g_2$ on the group manifold in a linear
fashion: $g(t) = g_1 + t (g_2 - g_1)~,~ 0 \leq t \leq 1$.

More explicitly, let $x$ and $x + a_g \mu$ be two neighboring
points on the coarse lattice, and
\begin{equation}
\label{lattice}
z_n = x + n a_{\rm det} \mu~,~~ n=0,1, \cdots N
\end{equation}
be the corresponding points on the fine lattice:
$z_0 = x, z_{N} = x + a_g \mu$, where $N = {a_g / a_{\det}}$. Then a
link
$$U_{x, \mu} = \exp \left[ i a_g \, G^m t^m \right]$$
is interpolated to a set of links as
\begin{equation}
\label{interpolate} \bar{U}_{z_n, z_{n+1}} = \exp \left[ i a_{\rm det} \, G^m
t^m \right]~~~,
\end{equation}
where the $t^m$ are SU(N) generators, and the bar denotes the
finer lattice. Equation (\ref{interpolate}) allows us to compute
$\bar{U}$ links which are sub-links of original $a_g$ link
variables, and lie on the outer perimeter of a plaquette. The
remaining $\bar{U}$ links, which are within an $a_g$ plaquette,
can be obtained through a similar interpolation starting from
opposite sides of the perimeter, yielding an entire set of
plaquettes on the $a_{\det}$ lattice. The field strengths
resulting from this interpolation are always of order $a_g^{-2}$
and can be kept small compared to the cutoff $\Lambda^2$. To
properly include the quasiparticle modes, the spacing of the
fermion lattice must be $a_{\det} \sim 1/ \mu$, while $a_g \sim 1/
\Lambda_{\rm QCD}$ is probably sufficient to capture the effects
of non-perturbative gauge configurations.

The fermion determinant is to be computed as a function of the
finer plaquettes $\{ \bar{U}_{x \mu} \}$. The result is
(approximately) real and positive and can be used for importance
sampling. Further, there is a physical understanding of the
complex part of the determinant: it originates in the modes far
from the Fermi surface which have been integrated out.

\section{Inequalities and Anomaly Matching}

Positivity of the measure allows for rigorous QCD inequalities at
asymptotic density. For example, inequalities among masses of
bound states can be obtained using bounds on bare quasiparticle
propagators. One subtlety that arises is that a quark mass term
does not lead to a quasiparticle gap (the mass term just shifts
the Fermi surface). Hence, for technical reasons the proof of
non-breaking of vector symmetries~\cite{Vafa:1984xg} must be
modified. (Naive application of the Vafa-Witten theorem would
preclude the breaking of baryon number that is observed in the
color-flavor-locked (CFL) phase~\cite{Alford:1998mk}). A
quasiparticle gap can be inserted by hand to regulate the bare
propagator, but it will explicitly violate baryon number. However,
following the logic of the Vafa-Witten proof, any symmetries which
are preserved by the regulator gap cannot be broken spontaneously.
One can, for example, still conclude that isospin symmetry is
never spontaneously broken (although see below for a related
subtlety). In the case of three flavors, one can introduce a
regulator $d$ with the color and flavor structure of the CFL gap
to show rigorously that none of the symmetries of the CFL phase
are broken at asymptotic density. On the other hand, by applying
anomaly matching conditions \cite{anomaly}, we can prove that the
$SU(3)_A$ symmetries {\it are} broken. We therefore conclude that
the CFL phase is the true ground state for three light flavors at
asymptotic density, a result that was first established by
explicit calculation
\cite{Evans:1999at,Hong:2000ru,Schafer:1999fe}.

To examine the long-distance behavior of the vector current,
we note that the correlator of the vector current
for a given gauge field $A$ can be written as
\begin{eqnarray}
\left<J_{\mu}^a(\vec v_F,x)J_{\nu}^b(\vec v_F,y)\right>^A =-{\rm
Tr}\,\gamma_{\mu}T^a S^A(x,y;d)\gamma_{\nu}T^b S^A(y,x;d),
\nonumber
\end{eqnarray}
where the $SU(N_f)$ flavor current $J_{\mu}^a(\vec v_F,x)
=\bar\psi_+(\vec v_F,x)\gamma_{\mu}T^a\psi_+(\vec v_F,x)$. The
propagator with $SU(3)_V$-invariant IR regulator $d$ is given as
\begin{eqnarray}
S^A(x,y;d)=\left<x\right|\frac{1}{M}\left|y\right>=\int_0^{\infty}
{\rm d}\tau \left<x\right|e^{-i\tau (-iM)}\left|y\right> \nonumber
\end{eqnarray}
where with $D=\partial+iA$
\begin{eqnarray}
M &=& \gamma_0
\begin{pmatrix}
D\cdot V \hfill & d \\
d^{\dagger} & D\cdot\bar V \hfill
\end{pmatrix}  \nonumber .
\end{eqnarray}
Since the eigenvalues of $M$ are bounded from below by $d$, we
have
\begin{eqnarray}
\left|\left<x\right|\frac{1}{M}\left|y\right>\right| \le
\int_R^{\infty}\!\!\!{\rm d}\tau \,e^{-d \,
\tau}\sqrt{\left<x|x\right>} \sqrt{\left<y|y\right>}=\frac{e^{-d
\, R}}{d} \sqrt{\left<x|x\right>}\sqrt{\left<y|y\right>},
\label{propagator1}
\end{eqnarray}
where $R\equiv\left|x-y\right|$.
The current correlators fall off rapidly as
$R\to \infty$;
\begin{eqnarray}
\left| \int {\rm d}A_+\!\!\!\right.\!\!\!\!\!& &\left.\!\!\!\!\! ~
\det M_{\rm eff}(A)\,\,e^{-S_{\rm eff}}
\left<J_{\mu}^A(\vec v_F,x)J_{\nu}^B(\vec v_F,y)\right>^{A_+}\right|\nonumber\\
&\le & \int_{A_+} \left| \left<J_{\mu}^A(\vec v_F,x)J_{\nu}^B(\vec v_F,y)
\right>^{A_+}\right| \le \frac{e^{-2d \, R}}{d^2}
\int_{A_+} \left|\left<x|x\right>\right|\left|\left<y|y\right>\right|,
\label{schwartz}
\end{eqnarray}
where we used the Schwartz inequality in the first inequality,
since the measure of the effective theory is now positive, and
equation (\ref{propagator1}) in the second inequality. The IR
regulated vector currents do not create massless modes out of the
vacuum or Fermi sea, which implies that there is no
Nambu-Goldstone mode in the $SU(3)_V$ channel. Therefore, for
three massless flavors $SU(3)_V$ has to be unbroken as in CFL. The
rigorous result provides a non-trivial check on explicit
calculations, and applies to any system in which the quasiparticle
dynamics have positive measure.

It is important to note the order of limits necessary to obtain
the above results. Because there are higher-order corrections to
the HDET, suppressed by powers of $\Lambda / \mu$, that spoil its
positivity, there may be contributions on the RHS of
(\ref{schwartz}) of the form
\begin{equation} \label{fR}
{\cal O} \left( \frac{\Lambda}{\mu} \right) ~f(R)~,
\end{equation}
where $f(R)$ falls off more slowly than the exponential in
(\ref{schwartz}). To obtain the desired result, we must first take
the limit $\mu \rightarrow \infty$ at fixed $\Lambda$ before
taking $R \rightarrow \infty$. Therefore, our results only apply
in the limit of asymptotic density.

Although our result precludes breaking of vector symmetries at
asymptotic density in the case of three {\it exactly} massless
quarks \cite{thermo}, it does not necessarily apply to the case
when the quark masses are allowed to be slightly non-zero. In that
case the results depend on precisely how the limits of zero quark
masses and asymptotic density are taken, as we discuss below.

In \cite{Bedaque:2001je} the authors investigate the effect of
quark masses on the CFL phase. These calculations are done in the
asymptotic limit, and are reliable for sufficiently small quark
masses. When $m_u = m_d \equiv m << m_s$ (unbroken $SU(2)$
isospin, but explicitly broken $SU(3)$), one finds a kaon
condensate. The critical value of $m_s$ at which the condensate
forms is $m_s^* \sim m^{1/3} \Delta_0^{2/3}$, where $\Delta_0$ is
the CFL gap (see, in particular, equation (8) of the first paper).
As kaons transform as a doublet under isospin, the vector $SU(2)$
symmetry is broken in seeming contradiction with our result.

However, a subtle order of limits is at work here. For simplicity,
let us set $m = 0$. Note that the CFL regulator $d$, which was
inserted by hand, explicitly breaks $SU(3)_A$ through color-flavor
locking, leading to small positive mass squared for the pions and
kaons, given as
\begin{eqnarray}
m_{\pi,K}^2 \sim \alpha_s d^2\,\ln\left(\frac{\mu}{d}\right).
\end{eqnarray}
The meson mass is not suppressed by $1/\mu$, since, unlike the
Dirac mass term, the regulator, being a Majorana mass, does not
involve antiquarks~\cite{Hong:1999ei}.

Therefore, even when the light quarks are massless, there is a
critical value of $m_s$ necessary to drive negative the
mass-squared of kaons and cause condensation:
\begin{equation}
m_s^* ~\sim~ \left[ g_s d \mu \, \ln\left(\frac{\mu}{d}\right)
\right]^{1/2} ~>~ ( d \mu )^{1/2}~,
\end{equation}
where $g_s$ is the strong coupling constant. Note the product of
$g_s$ with the logarithm grows as $\mu$ gets large. To obtain our
inequality we must keep the regulator $d$ non-zero until the end
of the calculation in order to see the exponential fall off. To
find the phase with kaon condensation identified
in~\cite{Bedaque:2001je} we must keep $m_s$ larger than $m_s^*$.
(Note $\mu \rightarrow \infty$, so to have any chance of finding
this phase we must take $d \rightarrow 0$ keeping $d R$ large and
$d \mu$ small.)

Since the UV cutoff of the HDET must be larger than $m_s$, we have
\begin{equation}
1 ~>~ \left( \frac{m_s^*}{\Lambda} \right)^2 ~>~ \frac{d}{\Lambda}
\, \frac{\mu}{\Lambda}~,
\end{equation}
which implies
\begin{equation}
\label{fR1}  \frac{\Lambda}{\mu} \, f(R) ~>~ \frac{d}{\Lambda} \,
f(R)~.
\end{equation}
Note the right hand side of this inequality does not necessarily
fall off at large $R$, and also does not go to zero for $\mu
\rightarrow \infty$ at fixed $\Lambda$ and $d$. This is a problem
since to apply our inequality the exponential falloff from
(\ref{schwartz}) must dominate the correction term (\ref{fR}),
which is just the left hand side of (\ref{fR1}). Combining these
equations, we see that the exponential falloff of the correlator
is bounded below,
\begin{equation}
\frac{e^{- 2d \, R}}{d^2} ~>~ {d \over \Lambda}  \, f(R)~,
\end{equation}
in the scaling region with a kaon condensate, $m_s > m_s^*$.

Alternatively, if we had taken $m_s$ to be finite for fixed
regulator $d$ (so that, as $\mu \rightarrow \infty$, eventually
$m_s < m_s^*$), the inequality in (\ref{schwartz}) could be
applied to exclude a Nambu-Goldstone boson, but we would find
ourselves in the phase without a kaon condensate.

\section{Conclusion}

The low-energy physics of dense fermionic matter, ranging from
quark matter to electronic systems, is controlled by modes near
the Fermi surface. An effective Lagrangian describing the
low-energy modes can be given in a systematic expansion in powers
of the energy scale over the chemical potential. The leading term
in this expansion has a simple form, and we have shown that it
leads to a real, positive Euclidean path integral measure.

This observation opens the door to importance sampling in Monte
Carlo simulations of dense matter systems. The key requirement is
that the interactions do not strongly couple the low-energy modes
to modes far from the Fermi surface. QCD at high density satisfies
this requirement, as do all asymptotically free models. Electronic
system in which the important interactions involve momentum
transfer less than the Fermi energy are in this category, although
some idealized models such as the Hubbard model are not. We have
given some proposals for how the positive effective theory might
be simulated numerically. Ultimately, we hope that actual
practitioners will develop even more practical methods.

Finally, positivity has analytical applications as well, since it
allows the use of rigorous inequalities. In QCD we obtain
restrictions on symmetry breaking patterns at high density.
Similar restrictions can probably be obtained for electronic systems
with suitable interactions.

\eject

\acknowledgments

We would like to thank M. Alford, S. Hands, C. Kim, W.
Lee, T.-S. Park, K. Rajagopal and M. Stephanov for useful
discussions. The work of D.K.H. was supported in part by
the academic research fund of Ministry of Education,
Republic of Korea, Project No. KRF-2000-015-DP0069.
The work of S.H. was supported in part under DOE contract
DE-FG06-85ER40224 and by the NSF through through the USA-Korea
Cooperative Science Program, 9982164.



\begin{thebibliography}{99}
\bibitem{vf}
For simplicity we set $v_F = 1$. Alternatively
it could be absorbed in the definition of the spatial $\gamma_1$
as in QCD below.
\bibitem{ev} As before, note that $\gamma_5$
anti-commutes with M, so if $M \phi = \lambda \phi$, then $M
\gamma_5 \phi = - \gamma_5 M \phi = - \lambda \gamma_5 \phi$.
\bibitem{gauge}
If we simultaneously gauge transform $A_+$ and $\psi_+$
in (\ref{leading2}) the result is invariant. There is a simple
relation between the gauge transform of the + fields and that of
the original fields: $U_+ (x) = U(x) e^{iX}.$ Of course, the
momentum-space support of the + gauge transform must be limited to
modes less than the cutoff $\Lambda$.
\bibitem{thermo} To investigate spontaneous symmetry breaking, one
ordinarily has to start at finite volume and insert a source which
explicitly breaks the symmetry. The source is removed only after
the infinite volume limit is taken. We stress that the source does
not have to be a quark mass (it could be a higher dimension
operator), so one can investigate symmetry breaking even when the
quark mass is exactly zero throughout the calculation. (To be
precise, a quark mass does not explicitly violate vector
symmetries, so it cannot play the role of the source in the
thermodynamic limit needed here.)
\end{thebibliography}
\end{document}